# Current-induced switching in single ferromagnetic layer nanopillar junctions

B. Özyilmaz[a] and A. D. Kent

Department of Physics, New York University, New York 10003, USA

Current induced magnetization dynamics in asymmetric Cu/Co/Cu single magnetic layer nanopillars has been studied experimentally at room temperature and in low magnetic fields applied perpendicular to the thin film plane. In sub-100 nm junctions produced using a nanostencil process a bistable state with two distinct resistance values is observed. Current sweeps at fixed applied fields reveal hysteretic and abrupt transitions between these two resistance states. The current induced resistance change is 0.5%, a factor of 5 greater than the anisotropic magnetoresistance (AMR) effect. We present an experimentally obtained low field phase diagram of current induced magnetization dynamics in single ferromagnetic layer pillar junctions.

75.60.Jk, 75.30.Ds, 75.75.1a

---

[a] Present Address: Department of Physics, Columbia University, New York, New York 10027, USA

In 1996 Slonczewski[1] and Berger[2] independently proposed a new mechanism by which the magnetization of a small ferromagnetic element may be manipulated when the latter is traversed by a spin polarized charge current. This ground breaking concept known as spin transfer, relies on the local interaction between a spin polarized current and the background magnetization of a ferromagnet. Spin transfer induced magnetization dynamics has been recently demonstrated experimentally in a wide variety of material systems and is rapidly reaching device maturity[3,4,5,6,7,8,9,10,11,12,13]. Furthermore, direct measurements of the magnetization dynamics[14,15,16] have shed light on the fundamental mechanism of spin transfer.

With a few exceptions[17,18] spin transfer devices consist of at least two exchange decoupled ferromagnetic regions, which play distinct roles. One of these regions is designed such that it simultaneously provides both a spin polarized current and acts as a reference layer relative to which the manipulation of the second ferromagnetic region is detected with the giant magnetoresistance effect (GMR). However, recently it has been demonstrated that such a distinction between a "fixed layer" and a "free layer" is not necessary. In large magnetic fields applied perpendicular to the thin film plane current induced reversible changes in junction resistance have been observed in pillar junctions containing only a single ferromagnetic layer[18]. These excitations have been associated with the onset of non-uniform spin wave modes[19,20,21,22]. Here the spin transfer mechanism relies on a feedback mechanism between the single layer magnetization and the spin accumulation in the adjacent nonmagnetic layers. Naturally the question arises whether similar physics takes place in the technologically relevant low field-low current regime.

In this letter we provide evidence that in magnetic fields below the demagnetization field spin transfer may lead to both reversible excitations and *hysteretic* changes in the junction resistance. Systematic, abrupt and hysteretic changes in device resistance have been observed in sub-100 nm size pillar junctions containing a single ferromagnetic (FM) layer in small perpendicular magnetic fields ($H<H_{demag}$). Both the threshold currents and the current induced hysteresis in these junctions exhibit a well defined magnetic field dependence. Similar to the high field excitations,[18] the low field single layer switching events lead to a low resistance state. However, the threshold currents and the corresponding changes in device resistance in the two regimes are very distinct. Remarkably, in the low field regime the magnetoresistance is comparable to that observed in the more traditional bilayer geometry. Thus, our findings may have important implications for the latter. For example, recent micromagnetic studies [23] predict considerable inhomogeneities of the free layer magnetization during magnetization reversal in bilayer junctions. Thus, it seems reasonable to assume that a feedback mechanism such as proposed here for the single layer junction may play an important role during the magnetization reversal in traditional bilayer junctions.

Asymmetric[b] pillar junctions ~ 50 nm in size with a stack sequence of |10 nm Pt|10 nm Cu| Co| 300 nm Cu| have been fabricated by means of a nano-stencil mask process (Fig. 1 inset). Many junctions with a Co layer thickness varying from 2 nm to 17

---

[b] Here asymmetry refers to the location of the Co layer within the junction. It is located much closer to the Pt layer.

nm and lateral dimensions from $30nm \times 60nm$ up to $70nm \times 210nm$ have been studied as a function of bias current and applied field. All junctions in this thickness range at magnetic fields $H < 4\pi M_s$ applied perpendicular to the thin film plane exhibit current induced hysteretic changes in the device resistance. Here we discuss representative data obtained on pillar junctions with ~ 15 nm thickness and lateral dimensions of $30nm \times 60nm$. Experiments were conducted at room temperature in a four point-geometry configuration in fields applied perpendicular to the thin film plane. The differential resistance *dV/dI* was measured by a lock-in technique with a 100 µA modulation current at f=873 Hz added to a dc bias current. Positive current is defined such that the electrons flow from the bottom electrode of the junction to the top electrode.

Characteristic current sweep traces for selected field values are shown in Fig. 1(a-c). Unipolar reversible dips similar to ones seen in large applied fields[18] ($H >> 4\pi M_s$) are observed in fields down to *H < 0.6 T*. In smaller fields, reversible dips are observed for *both* current polarities. There is a dramatic change in the current sweep traces when the magnetic fields are lowered further. At fields below *H < 0.5 T* we observe hysteretic and abrupt changes in the junction resistance. We discuss the current induced switching event at 250 mT in more detail. Starting from +9 mA the junction remains in a high resistance state until $I \approx -8 mA$, where it switches to a resistance state that has a 0.5% smaller resistance. The junction remains in the low resistance state till the current sweep direction is reversed and reaches +1.5 mA. At this current value the junction switches back to a high resistance state. Note that in both the switching from a low resistance state to a high resistance state and vice versa the junction resistance changes by ~ 0.5%. It is intriguing to compare this resistance change with the anisotropic magnetoresistance (AMR) at zero dc bias of similar junctions. The latter was measured in such junctions at 4.2 K and was found to be ~ 0.1%[18, 24]. The current induced hysteretic resistance change at low magnetic fields and room temperature is a factor of 5 larger and thus cannot be associated with the AMR effect.

The applied field dependence of current induced single layer switching (CISS) is best summarized by plotting the differential resistance $dV/dI$ during the current sweep up (Fig. 2(a)) and current sweep down (Fig. 2(b)) measurements on separate color plots. To clearly distinguish CISS from the varying background resistance, we plot the differential resistance minus a polynomial fit. Here the current is swept from +9 mA to -9 mA and back to +9 mA while the magnetic field is held constant. In subsequent current sweeps the applied field is reduced to zero from 0.9 T in steps of 5 mT. From Fig. 2 we see that the critical currents necessary to switch the junction resistance to a low resistance state shift to more negative currents with decreasing applied fields such that below 0.21 T the junction resistance remains in the high resistance state. On the other hand, the width of the current induced hysteresis increases with decreasing applied field and at 0.21 T it reaches approximately 10 mA. Note that the current induced change in junction resistance depends on both the applied field and the current bias; $\Delta R/R$ decreases strongly with increasing applied field and weakly with increasing current bias.

Even in single layer junctions the fundamental mechanism for current induced excitations remains spin transfer as proposed by Slonczewski[1]. Here fluctuations of the magnetization in conjunction with the diffusive transport at the normal metal-ferromagnet interfaces are believed to cause a locally transverse spin accumulation and hence generate spin transfer torques. In fields $H < 4\pi M_s$ non-uniformities in the magnetization could provide directly for a local transverse spin accumulation along the interface. We note that current induced switching events are also observed in the field in-plane configuration[24]. However, neither the switching threshold nor the hysteresis width varies systematically with magnetic field. The absence of systematic current induced switching events in the in-plane geometry shows that the observed hysteresis is not caused by the presence of an extended ferromagnetic layer within the top electrode. This layer has a minimal impact both because it is well separated from the nanomagnet (see Fig. 1, inset) and there is an intervening Pt layer, which is a strong spin scatter. In fact, we find that the MR observed in in-plane field sweeps at zero dc bias is an order of magnitude smaller than the current induced MR.

One possible magnetic configuration which could provide for both a bistable state in the field perpendicular geometry and a distinct dependence on the applied field direction is a vortex state. In a perpendicular field below the demagnetization field, a change of the applied field, leads to a change of the angle between the neighboring magnetic moments of the disk magnetization and between the disc magnetization and the applied field itself. Thus, it seems reasonable to assume that here the spin transfer torques show a strong magnetic field dependence. Evidently transport measurements are not well suited for deducing the nature of the two magnetic states. However, the large decrease/change in junction resistance does provide evidence that the two resistance states correspond to markedly distinct magnetic configurations. Recently, it has been argued that a non-uniform magnetization can lower the junction resistance[18, 21, 25]. In contrast to a uniform magnetization, a non-uniform magnetization provides a mechanism for the mixing of the spin channels and thus decreases the net spin accumulation across the pillar junction. Again, a vortex state would be a good candidate for the low resistance state. An ideal candidate for a high resistance state would be a single domain ferromagnetic layer. Thus one possible scenario for the observed hysteretic change in junction resistance could be a current induced change of the layer magnetization from a vortex state to a state of near uniform magnetization.

To summarize our results give indirect evidence for an "intrinsic" non-uniform magnetization distribution in sub 100 nm junction within single ferromagnet layer pillar junctions and demonstrate that even at low current bias, feedback effects between the layer magnetization and the spin accumulation in the adjacent nonmagnetic layers may have to be taken into account to fully understand spin transfer. Our results show that the magnetoresistance in a single layer junction can be large and cannot be associated with the AMR effect. The mechanism is likely associated with a change in spin accumulation. This may provide an alternative way to read out the magnetic state of a pillar junction. It seems reasonable to assume that these effects may play also an important role in the magnetization dynamics of spin transfer devices consisting of more than one ferromagnetic layer. Our findings may help to optimize the performance spin transfer devices.

Acknowledgements: This research was supported by NSF-DMR-0405620. We thank Jonathan Z. Sun and Michael J. Rooks for useful comments and fabrication of the nanopillar junctions used in this study.

**Figure1: Current sweeps at fixed applied fields. (a) At high fields reversible dips in dV/dI are only observed at negative currents. (b) At intermediate field values dips in dV/dI are observed for both current polarities. (c) At low magnetic fields current sweeps exhibit hysteretic behavior. Inset shows current polarity convention and layer structure of a nanostencil produced single layer junction.**

**Figure 2: Low field phase diagram for current induced excitations in single layer pillar junctions. The differential resistance minus a polynomial fit is plotted on a color scale as a function of both applied field and current bias. White solid arrows indicate current sweep direction and dashed arrows indicate field step direction. (a) Current sweep. (b) Current sweep down. The region where hysteresis is observed is indicated by the black dashed dotted line.**

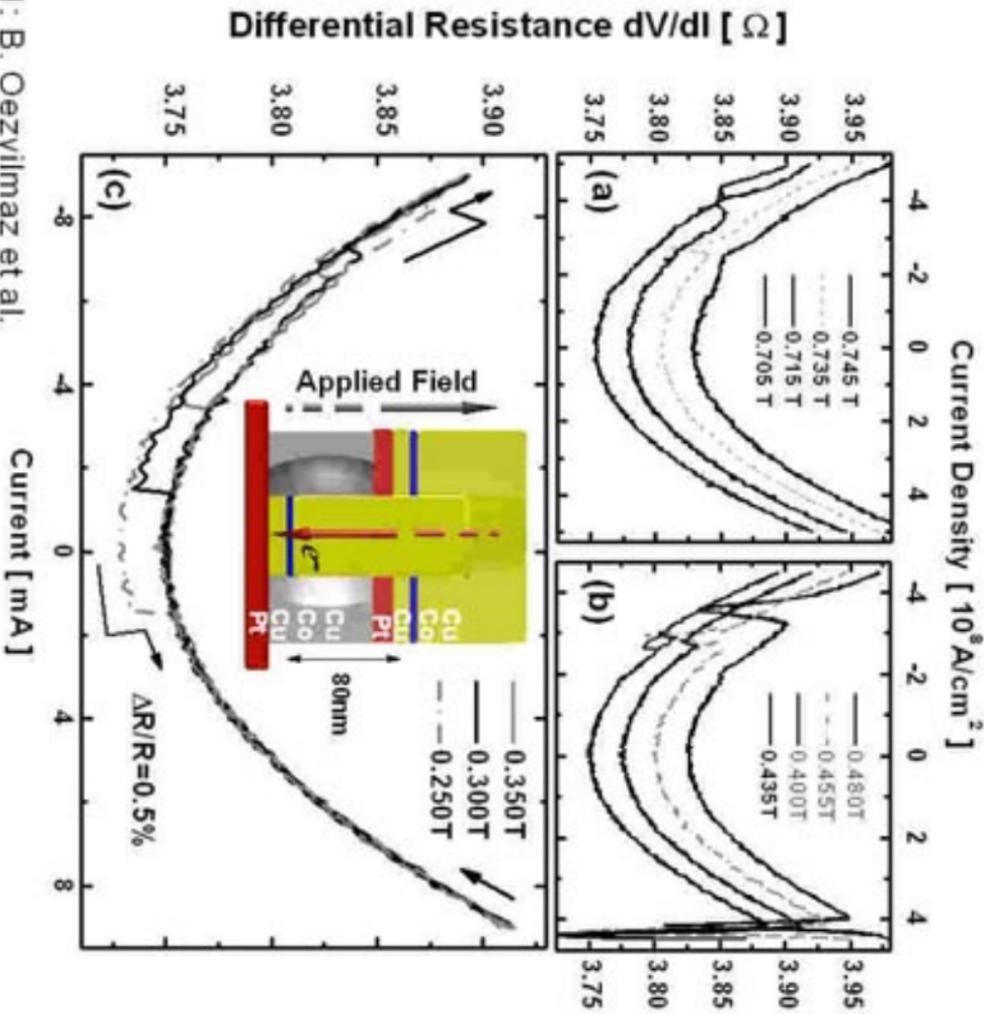

Fig. 1: B. Oezyilmaz et al.

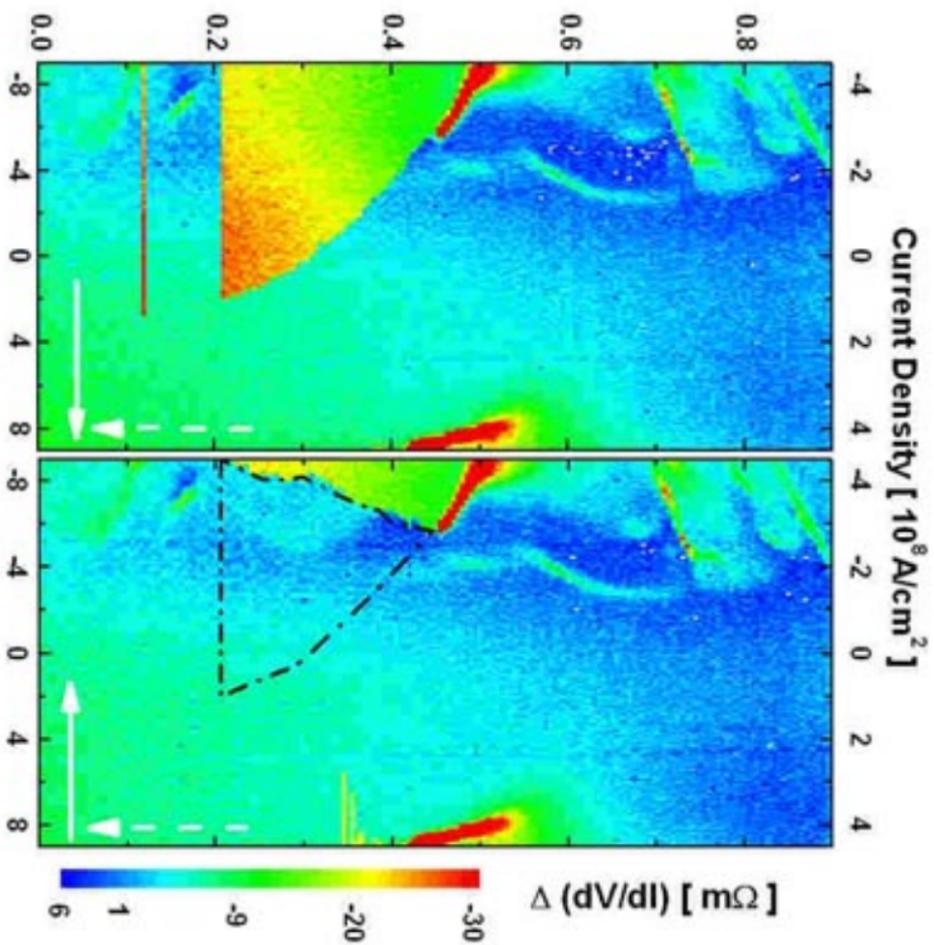

Fig. 2: B. Oezyilmaz et al.